\documentclass[12pt]{article}
\begin{document}

\begin{titlepage}

\begin{flushright}
hep-th/0004138
\\
FTUV-0418,\ IFIC/00-25
\\
MPI-PhT 2000-14
\\
NYU-TH/00/03/07
\\
ULB-TH-00/12
\end{flushright}

\begin{centering}


\huge{Gauge dependence of effective action and renormalization group
functions in effective gauge theories} \\

\vspace{.5cm}

\large{G. Barnich$^*$}\\

Physique Th\'eorique et Math\'ematique, Universit\'e Libre de Bruxelles,
Boulevard du Triomphe, Campus Plaine C.P. 231, B-1050
Bruxelles, Belgium

\vspace{.25cm}

and\\

\vspace{.25cm}

\large{P.A. Grassi}\\
Department of Physics, New York University,  \\
4 Washington Place, New York, NY 10003, USA

\end{centering}

\vspace{.5cm}

\begin{abstract}
The Caswell-Wilczek analysis on the gauge dependence of the effective
action and the renormalization group functions in Yang-Mills theories
is generalized to generic, possibly power counting non renormalizable
gauge theories.  It is shown that the physical coupling constants of the
classical theory can be redefined by gauge parameter dependent
contributions of higher orders in $\hbar$ in such a way that
the effective action depends trivially on the gauge parameters, while
suitably defined physical beta functions do not depend on those
parameters.
\end{abstract}


\footnotesize{$^*$ Scientific Research Worker of the
FNRS (Belgium).}

\end{titlepage}

\addtolength{\topmargin}{-2cm}
\addtolength{\textheight}{3.5cm}
\addtolength{\oddsidemargin}{-1cm}
\addtolength{\textwidth}{1.5cm}
\addtolength{\footskip}{0.7cm}
\sloppy

\def\bea{\begin{eqnarray}}
\def\eea{\end{eqnarray}}
\def\beann{\begin{eqnarray*}}
\def\eeann{\end{eqnarray*}}
\def\beq{\begin{equation}}
\def\eeq{\end{equation}}
\def\ba{\begin{array}}
\def\ea{\end{array}}
\def\ben{\begin{enumerate}}
\def\een{\end{enumerate}}
\def\bea{\begin{eqnarray}}
\def\eea{\end{eqnarray}}
\def\beann{\begin{eqnarray*}}
\def\eeann{\end{eqnarray*}}
\def\beq{\begin{equation}}
\def\eeq{\end{equation}}
\def\ba{\begin{array}}
\def\ea{\end{array}}
\def\ben{\begin{enumerate}}
\def\een{\end{enumerate}}

\def\5{\bar }
\def\6{\partial }
\def\7{\hat }
\def\4{\tilde }

\def\gh{\mbox{gh}}
\def\agh{\mbox{antigh}}
\def\tot{\mbox{totdeg}}
\def\deg{\mbox{formdeg}}

\def\cA{{\cal A}}
\def\cB{{\cal B}}
\def\cC{{\cal C}}
\def\cD{{\cal D}}
\def\cE{{\cal E}}
\def\cF{{\cal F}}
\def\cG{{\cal G}}
\def\cH{{\cal H}}
\def\cI{{\cal I}}
\def\cJ{{\cal J}}
\def\cK{{\cal K}}
\def\cL{{\cal L}}
\def\cM{{\cal M}}
\def\cN{{\cal N}}
\def\cO{{\cal O}}
\def\cP{{\cal P}}
\def\cQ{{\cal Q}}
\def\cR{{\cal R}}
\def\cS{{\cal S}}
\def\cT{{\cal T}}
\def\cU{{\cal U}}
\def\cV{{\cal V}}
\def\cW{{\cal W}}
\def\cX{{\cal X}}
\def\cY{{\cal Y}}
\def\cZ{{\cal Z}}

\def\s0#1#2{\mbox{\small{$\frac{#1}{#2}$}}}
\def\chris#1#2#3{{\Gamma_{#1#2}}^{#3}}
\def\f#1#2#3{{f_{#1#2}}^{#3}}
\def\fd#1#2#3{{f_{#1#2#3}}}
\def\A#1#2{{A_{#2}}^{#1}}
\def\viel#1#2{e_{#2}{}^{#1}}
\def\Viel#1#2{E_{#1}{}^{#2}}
\def\csum#1#2{\sum_{#1}\hspace{-1.#2em}\circ\ \ \ }
\def\G{\Gamma}
\def\D{\Delta}

\newtheorem{theorem}{Theorem}
\def\qed{\hbox{${\vcenter{\vbox{
\hrule height 0.4pt\hbox{\vrule width 0.4pt height 6pt
\kern5pt\vrule width 0.4pt}\hrule height 0.4pt}}}$}}

\newtheorem{lemma}{Lemma}

\section{Introduction}
The problem of the gauge dependence of the effective action and of
the renormalization group functions has been extensively studied
in the mid seventies in the context of Yang-Mills theories
\cite{Caswell:1974cj,Kluberg-Stern:1975rs,Kluberg-Stern:1975xv,
Kluberg-Stern:1975hc,Breitenlohner:1975qe}. An algebraic approach to
the problem, independent of the renormalization scheme, has been
proposed in \cite{Piguet:1985js}. On the assumption of the
existence of an invariant renormalization scheme, extensions to
generic, not necessarily power counting renormalizable theories have
been considered in
\cite{Voronov:1982cp,Voronov:1982ur,Voronov:1982ph,Lavrov:1985hr}
and more recently in \cite{Anselmi:1994ry,Anselmi:1995zx}.

In this letter, we combine the ideas of the above cited works and
reinvestigate the problem in a general setting. We clarify the
essential points of the analysis by getting rid of unnecessary
simplifying assumptions. More precisely:
\begin{itemize}
\item
The analysis covers effective
theories, i.e., theories that are not necessarily
assumed to be power counting renormalizable. An example
is Yang-Mills theory (based for simplicity on a
semi-simple gauge group) involving higher dimensional gauge invariant
operators as considered in
\cite{Gomis:1996jp}.
\item The considerations are not restricted to Yang-Mills type theories, but
    they extend to the case of generic reducible gauge theories with structure
functions and open algebras \cite{Batalin:1981jr,Batalin:1983jr}.
\item The particular way the gauge is fixed is irrelevant. In
    particular, we do not need to restrict ourselves to the case of
linear gauges.
\item We do not assume the existence of a gauge invariant renormalization
    scheme.
\end{itemize}

In order to control the renormalization
aspects of the problem, independently of the particular scheme being used, we
assume that the quantum action principles
\cite{Lowenstein:1971jk,Lowenstein:1971vf,Lam:1972mb,Lam:1973qa,Clark:1976ym}
hold and follow the algebraic approach pioneered in
\cite{Lowenstein:1973qt,Becchi:1974xu,Becchi:1975md,Becchi:1976nq}
   (for reviews, see e.g. \cite{Piguet:1981nr,Piguet:1995er}).

\section{Preliminaries}

\subsection{Master equation and gauge fixing}
\label{prelim}

Gauge invariance of the classical action $S_0[\varphi^i]$
and the algebra of the
gauge transformations are encoded in the minimal solution
$S[\phi^A,\phi^*_A]$ of the master equation
\cite{Zinn-Justin:1974mc,Zinn-Justin:1989mi,Batalin:1981jr,Batalin:1983jr}
(for reviews, see e.g.
\cite{Henneaux:1992ig,Gomis:1995he}):
\beq
\frac{1}{2}(S,S)_{\phi,\phi^*}=0.\label{ma}
\eeq
The gauge fixing can be done in two
steps: first one adds a cohomological trivial non minimal
sector. This amounts to extending the
minimal solution of the master equation to $S^\prime=S+\int d^nx\
B^a\bar C^*_a$.
The canonical BRST differential
extended to the antifields and the non minimal sector is
$s=(S^\prime,\cdot)_{\phi,\phi^*}$.
The second step is to perform an anticanonical transformation generated
by a gauge fixing fermion $\Psi[\phi^A]$: the gauge fixed action to
be used for quantization is $S_{\rm gf}[\phi^A,\tilde\phi^*_A]=
S^\prime[\phi^A,\tilde\phi^*_A+\frac{\delta^L\Psi}{\delta
    \phi^A}]$, with $\Psi$ chosen in such a way that the propagators
of the theory are well defined.
For instance, in Yang-Mills type theories, standard linear gauges are
obtained from
\bea
\Psi=\int d^nx\ \bar
C_a(\partial^\mu A^a_\mu +\frac{1}{2}\xi B^a).\label{psi}
\eea
The  cohomology of the associated
BRST differential $s=(S_{\rm gf},\cdot)_{\phi,\tilde\phi^*}$ in the space of
local functions or in the space of local functionals is
isomorphic to the cohomology of the canonical BRST differential in the
respective spaces and can be
obtained from it through the shift of antifields
$\phi^*=\tilde\phi^*+\delta^L\Psi/\delta\phi$.
The dependence of the gauge fixed action on the fields and
antifields of the non
minimal sector and their gauge fixed BRST transformations are
explicitly given by
\bea
s\, \tilde B^*_a = \frac{\delta^R S_{\rm gf}}{\delta B^a} = -(S_{\rm
gf},\frac{\delta^R\Psi}{\delta B^a})_{\phi,\tilde\phi^*}+\tilde{\bar C^*_a},
~~~~
s\, B_a = -\frac{\delta^R
S_{\rm gf}}{\delta\tilde{B^*_a}}=0,\label{5bis}
\eea
\bea
s\, \tilde{\bar C^*_a}= \frac{\delta^R
S_{\rm gf}}{\delta\bar C^a}=-(S_{\rm
gf},\frac{\delta^R\Psi}{\delta\bar C^a})_{\phi,\tilde\phi^*},
~~~~
s\, \bar C^a = -\frac{\delta^R
S_{\rm gf}}{\delta\tilde{\bar C^*_a}}=-B^a.
\eea
These transformations are nilpotent and guarantee that the BRST
cohomology does not depend on the  fields and the antifields of the
non minimal sector.

The renormalized effective action associated to the gauge fixed action
$S_{\rm gf}[\phi,\tilde\phi^*]$ is denoted by $\Gamma_{\rm
gf}[\phi_c,\tilde\phi^*]$.

\subsection{Assumptions on anomalies and couplings}

Throughout the analysis, we make the following assumptions:
\begin{itemize}
\item The theory is stable in the sense that the BRST cohomology in
ghost number $0$ in the space of appropriate
local functionals can be obtained by
differentiation with respect to some couplings of the minimal solution  of
the master equation:
\beq
(S,A)_{\phi,\phi^*}=0, {\rm gh}\ A=0\Longrightarrow A=
\lambda^i\partial_{g^i}S+(S,\Xi)_{\phi,\phi^*}.\label{coh}
\eeq
\item The
gauge symmetry is non anomalous in the sense that the
Zinn-Justin equation
\beq
\frac{1}{2}(\Gamma_{\rm gf},\Gamma_{\rm gf})_{\phi_c,\tilde\phi^*}=0,\label{-1}
\eeq
holds, by adding (if necessary) finite BRST breaking
counterterms to the starting point action. This is the case if the
BRST cohomology in the space of local functionals in ghost number $1$
is empty, or if one can prove that the corresponding anomaly candidates
do not effectively arise because their coefficients vanish to
all orders in $\hbar$.
\item The couplings $g^i$ are non redundant in the sense that
\beq
\mu^i\partial_{g^i} S
=(S,\Xi^\prime)_{\phi,\phi^*}\Longrightarrow \mu^i=0
=(S,\Xi^\prime)_{\phi,\phi^*}.\label{red}
\eeq
As a consequence of this definition the value of the
non-redundant couplings is fixed in terms of observables. This
procedure automatically ensures that there is no mixing among
physical and unphysical (or
redundant) couplings (see for example \cite{Gambino:1999ai}).
However, it implies that one should rely on a specific renormalization scheme.
In the following we will show that it is indeed possible to obtain
gauge-parameter independent
quantities without using a specific renormalization scheme.

\end{itemize}
{\bf Remarks:}

(i) Because the gauge fixing is an anticanonical transformation, the
relations (\ref{coh}) and (\ref{red})
hold in terms of $S_{\rm gf}$, local functionals
$\Xi,\Xi^\prime$ modified through the replacement
$\phi^*\longrightarrow \tilde\phi^*+
\delta\Psi/\delta\phi$ and the antibrackets in terms of
$\phi,\tilde\phi^*$.

(ii) What the appropriate space of local functionals is precisely
depends on the context. Usually it is the space of
integrals of $x^\mu$ independent polynomials or power series in the
couplings, the $dx^\mu$, the fields, antifields and their derivatives,
which can be further restricted by global symmetries such as Lorentz
invariance or by power counting arguments. In particular, in the case of
theories with massless and massive particle, the presence of IR singularities
might restrict the space of local functionals. However, this situation
can be handled by defining
a proper IR power counting \cite{Piguet:1995er}.

(iii) The last assumptions means that the coupling constants $g^i$ are
associated to independent BRST cohomological classes. It is the crucial
assumption that allows to extend the Caswell-Wilczek arguments to
generic gauge theories.

We assume that the $g^i$ are the only couplings on which the minimal
solution of the
master equation depends. They can be considered
as the ``physical couplings on the classical level''.
Note that in the gauge
fixed theory, for any parameter $\xi^\alpha$
appearing in $\Psi$, we have
\beq
\partial_{\xi^\alpha} S_{\rm gf}=-(S_{\rm gf},
\partial_{\xi^\alpha}\Psi)_{\phi,\tilde\phi^*},\label{5}
\eeq
which means in particular that all the parameters introduced through
the gauge fixing alone are redundant.

Hence, we will assume that in the gauge fixed theory,
the only additional couplings besides the physical $g^i$'s,
are the redundant gauge couplings $\xi^\alpha$ satisfying (\ref{5}).
Notice that the wave function normalization constants are redundant
couplings, since they can be introduced through anticanonical field
antifield redefinitions.

\section{``Physical'' coupling constants on the quantum level}

According to the quantum action principle, $\partial_{\xi^\alpha}
\Gamma_{\rm gf}=K_\alpha\circ\Gamma_{\rm gf}$, where $K_\alpha=
-(S_{\rm gf},\partial_{\xi^\alpha}\Psi)_{\phi_c,\tilde\phi^*}
+O(\hbar)$. It follows from lemma \ref{00} of the appendix that
this implies in a first step
\beq
\partial_{\xi^\alpha}
\Gamma_{\rm gf}=(\Gamma_{\rm gf},[-\partial_{\xi^\alpha}\Psi]^Q\circ
\Gamma_{\rm gf})_{\phi_c,\tilde\phi^*}
+\hbar K^\prime_\alpha\circ\Gamma_{\rm gf}.\label{-6}
\eeq
Here $[-\partial_{\xi^\alpha}\Psi]^Q$ is the renormalized operator
$\partial_{\xi^\alpha}\Psi$. Notice that it requires a renormalization which
independent from the renormalizations needed for the effective action $\Gamma$.
In the literature
\cite{Lowenstein:1972pr,Bandelloni:1980sp,Piguet:1985js}, different
approaches
have been used to define  $[-\partial_{\xi^\alpha}\Psi]^Q$ based on
the Wilson expansion or on the
extended BRST technique \cite{Piguet:1985js}. All of these approaches
amount to obtain the equation
(\ref{-6}) where $K^\prime_\alpha$ can be studied algebraically.

One can then go on to show (see also the appendix) that
\beq
\Big[\partial_{\xi^\alpha}
+\hbar\rho^i_\alpha\partial_{g^i}+
(L_{\alpha}\circ
\Gamma_{\rm gf},\cdot)_{\phi_c,\tilde\phi^*}\Big]\Gamma_{\rm gf}=0,\label{2.1}
\eeq
where $L_{\alpha}=-\partial_{\xi^\alpha}\Psi+O(\hbar)$ and the coefficients
$\rho^i_\alpha$ are formal power series in $\hbar$ depending on the
couplings $g^i$ and $\xi^\beta$.

Let us define $D_\alpha=\partial_{\xi^\alpha}+
\hbar\rho^i_\alpha\partial_{g^i}$.
By adapting the extended BRST technique of \cite{Piguet:1985js} to the
present context, one can show (see appendix) that there exist
local functionals $K_{[\alpha\beta]}$ such that
\beq
[D_{\alpha},D_{\beta}]^i\partial_{g^i}\Gamma_{\rm gf}=
-(K_{[\alpha\beta]}\circ
\Gamma_{\rm gf},\Gamma_{\rm gf})_{\phi_c,\tilde\phi^*}.\label{2.2}
\eeq
It then follows from the lemma \ref{l1} proved in the appendix that
\beq
[D_{\alpha},D_{\beta}]^i=0=(K_{[\alpha\beta]}\circ
\Gamma_{\rm gf},\Gamma_{\rm gf})_{\phi_c,\tilde\phi^*}.
\eeq
$[D_{\alpha},D_{\beta}]^i=0$ reads explicitly
$\partial_{\xi^[\beta}\rho^i_{\alpha]}
+\hbar\rho^j_{[\beta}\partial_{g^j}\rho^i_{\alpha]}=0$, which gives
to lowest order in $\hbar$ the relation
$\partial_{\xi^{[\beta}}\rho^i_{0\alpha]}=0$. Using the standard
Poincar\'e lemma (assuming that the gauge parameters space has trivial
topology), there exist functions $G^i_1(\xi,g)$ such that
$\rho^i_{0\alpha}=\partial_{\xi^\alpha}G^i_1(\xi,g)$.
Let us now define new couplings $g^i_1=g^i-\hbar G^i_1(\xi,g)$ and the
inverse transformation $g^i=g^i_1+\hbar G^i_1(\xi,g_1)+O(\hbar^2)$.
An explicit integration formula for $G^i_1(\xi,g)$ has been
given in refs.~\cite{Lowenstein:1972pr,Breitenlohner:1975qe}.
 
If we denote the generating functional in terms of the new couplings
with a subscript $1$, $\Gamma_{1{\rm gf}}(g_1,\xi)
=\Gamma_{{\rm gf}}(g(g_1,\xi),\xi)$ and
use the same notation for all functionals,
it follows that
$
\partial_{\xi^\alpha} \Gamma_{1{\rm gf}}+((L_{1
\alpha}\circ\Gamma_{1{\rm gf}}
,\Gamma_{1{\rm gf}})_{\phi_c,\tilde\phi^*}
+\hbar^2\bar \rho^i_\alpha(g_1,\xi)\partial_{g^i_1}\Gamma_{1{\rm gf}}=0,$
for some $\bar \rho^i_{\alpha}(g_1,\xi)$.
By a succession of redefinitions of the couplings $g^i$, we can thus
achieve (dropping the subscripts)
\beq
\partial_{\xi^\alpha} \Gamma_{\rm gf}+(L_\alpha\circ \Gamma_{\rm gf},
\Gamma_{\rm gf})_{\phi_c,\tilde\phi^*}\label{6}
=0.
\eeq
This leads to the following definition:
 
{\em The physical coupling constants $g^i$ on the quantum level
are such that the variation of the
effective action with respect to
the gauge parameters is given by
$(\Gamma_{\rm gf},\cdot)$ acting on a local insertion.}

It is the natural generalization of what one considers as physical on the
classical level. It follows that physical couplings $g^i$ on the classical
level stay physical in the quantum theory, by using the additional
freedom of redefinitions of the $g^i$ by terms of higher order in
$\hbar$ involving the gauge parameters.

After projection on the physical states, equation (\ref{6}) implies
the gauge parameter independence of $\Gamma_{\rm gf}$. Together
with the BRST invariance expressed through the Zinn-Justin equation
(\ref{-1}), these equations are the substitute for the gauge
invariance of the original action. A kind of direct gauge invariance
for  $\Gamma_{\rm gf}$ can be achieved using the background field method,
which will not be discussed here.

The procedure presented here differs from the conventional approach
to gauge-parameter
independent quantities (see for example \cite{Gambino:1999ai}) since
it does not rely on a specific
renormalization scheme and on the physical observable used to fix the
renormalization constants.
A similar approach has been pursued in
\cite{Breitenlohner:1975qe,Piguet:1989pc} following the work of
Zimmermann \cite{Zimmermann:1985sx}.

\section{Physical beta functions in the renormalization group equation}

Let us start for simplicity with the case where
the theory is renormalizable by constant redefinitions of
the fields and the antifields and by coupling constant redefinitions.
Then, the renormalization group equation is
\beq
[\mu\partial_\mu+\hbar\beta^i\partial_{g^i}
+\hbar\delta^\alpha\partial_{\xi^\alpha}]
\G_{\rm gf}+\hbar {\gamma^A_B}(\int d^nx\  \tilde\phi^*_A\phi^B_c,
\G_{\rm gf})_{\phi_c,\tilde\phi^*}=0.\label{1}
\eeq
As in \cite{Caswell:1974cj}, we inject (\ref{2.1}) respectively
(\ref{6}) into (\ref{1}) to get
\beq
[\mu\partial_\mu+\hbar\bar\beta^i\partial_{g^i}]
\G_{\rm gf}+\hbar(C\circ\Gamma_{\rm gf},\G_{\rm gf})_{\phi_c,\tilde\phi^*}=0,
\label{2}
\eeq
where
$\bar \beta^i=\beta^i-\hbar\delta^\alpha\rho^i_\alpha$ and $C=
\int d^nx\ \gamma^{A}_B  \tilde\phi^*_A\phi^B_c-\hbar\delta^\alpha
L_\alpha$.
Note that in the second case $\bar
\beta^i=\beta^i$, because $\rho^i_\alpha=0$.

In the general case,
it is still possible to prove directly that (\ref{2}) holds,
for some $\bar\beta^i$ and some local insertion $C\circ\G_{\rm gf}$ (see
appendix). This leads to the following definition:

{\em The physical beta functions $\bar \beta_i$ of the renormalization
group equation
are the coefficients of the derivatives $\partial_{g^i}$ associated to
physical couplings $g^ i$ of the classical level,
in the renormalization group equation where
the derivatives with respect to the redundant couplings have been eliminated.}

The derivation of (\ref{2}) given in the appendix shows that it is
always possible to cast the renormalization group equation in this
form as long as the quantum action principle holds, and the theory is non
anomalous and stable.

If one follows \cite{Caswell:1974cj} and
commutes the functional operators
of equations (\ref{2.1}) and (\ref{2}), one gets by defining $D=
   \mu\partial_\mu+\hbar\bar \beta^j\partial_{g^j}$,
\beq
[D,D_\alpha]^i\partial_{g^i}+(E_\alpha,\Gamma_{\rm
gf})_{\phi_c,\tilde\phi^*}=0,
\eeq
where
$E_\alpha=D[L_\alpha\circ\Gamma_{\rm gf}]-D_\alpha[\hbar C\circ\Gamma_{\rm
gf}]+(\hbar C\circ\Gamma_{\rm
gf},L_\alpha\circ\Gamma_{\rm gf})_{\phi_c,\tilde\phi^*}
$.

Again, we deduce $[D,D_\alpha]^i=0$ (see appendix).
If one uses the physical couplings of the quantum level, where
$\rho^i_\alpha=0$, these relations reduce to
\beq
\partial_{\xi^\alpha}\bar\beta^i=0.
\eeq
This gives the main result:

{\em In a non anomalous stable theory, the
physical $\beta$ functions do not depend on the
gauge parameters of the theory, if the effective action is
expressed in terms of physical coupling constants of the quantum
level.}

We also note that if one integrates the renormalization group equation
\beq
\mu\frac{d}{d\mu} G^i(g,\mu)=\hbar\bar\beta^i(G,\mu)
\eeq
and replaces the
couplings $g^i$ by the running couplings $G^i(g,\mu)$ in the effective
action,  equation (\ref{2}) reduces to the statement that the
renormalization scale
dependence of the effective action is given by $(\Gamma_{\rm
  gf},\cdot)$ acting on a local insertion,
\beq
\mu\frac{d}{d\mu} \G_{\rm gf}
+\hbar(C\circ\Gamma_{\rm gf},\G_{\rm gf})_{\phi_c,\tilde\phi^*}=0.
\eeq
Again, after projection on the physical states, this equation
expresses the renormalization scale independence of the effective
action. The compatibility conditions $[D,D_\alpha]^i=0$ and $[D_\alpha
,D_\beta]^i=0$ guarantee that the various redefinitions of the
couplings $g^i$ can be done simultaneously. 

\section{``Physical'' effective action}

Let us define the functionals
\beq
\Gamma^\prime[\phi_c,\phi^*_c]=\Gamma_{\rm gf}[\phi_c,\phi^*_c-\frac{\delta
\Psi(\phi_c)}{\delta\phi_c}],
\eeq
and
\beq
\Gamma[\phi_c,\phi^*_c]=\Gamma^\prime[\phi_c,\phi^*_c]-\int d^nx\
B^a_c\bar {C_a}_c.
\eeq
In other words, we undo, after quantization, the gauge fixing on the
level of the effective action.
The gauge fixing procedure and the passage to
$\Gamma$ can be summarized by the following diagram:
\beann
\begin{array}{r}
S[\phi,\phi^*]\rightarrow S^\prime[\phi,\phi^*]=
S[\phi,\phi^*]+\int d^nx B^a\bar {C^*_a} \rightarrow
S_{\rm gf}[\phi,\tilde\phi^*]=
S^\prime(\phi,\tilde\phi^*+\frac{\delta\Psi}{\delta \phi})\\
    \downarrow\ \ \ \ \  \\
\G[\phi_c,\phi^*_c]=\G^\prime[\phi_c,\phi^*_c]
-\int d^nx B^a_c\bar {C^*_a}_c\leftarrow
\G^\prime[\phi_c,\phi^*_c]=\Gamma_{\rm gf}
[\phi_c,\phi^*_c-\frac{\delta\Psi}{\delta \phi_c}]
\leftarrow
\Gamma_{\rm gf}[\phi_c,\tilde\phi^*]
\end{array}
\eeann
Because the shift in the antifields is a canonical transformation,
(\ref{-1}) implies
\beq
\frac{1}{2}(\G^\prime,\G^\prime)_{\phi_c,\phi^*_c}=0.\label{0}
\eeq
Furthermore,
$\G^\prime[\phi_c,\phi^*_c]=S^\prime[\phi_c,\phi^*_c]+O(\hbar)$ and
$\G[\phi_c,\phi^*_c]=S[\phi_c,\phi^*_c]+O(\hbar)$.
Note however that  $\Gamma^\prime[\phi_c,\phi^*_c]$ or
$\Gamma[\phi_c,\phi^*_c]$ cannot be interpretated
directly as the generating functional for 1PI vertex functions
associated to $S^\prime[\phi,\phi^*]$, respectively $S[\phi,\phi^*]$,
since these actions are gauge invariant and cannot be used to derive
Feynman rules. Rather, a particular Green's functions of
$\Gamma[\phi_c,\phi^*_c]$ is
given by the combination of 1PI vertices
of $\Gamma_{\rm gf}[\phi^A_c,\tilde\phi^*_A]$ obtained by using the
chain rule of differentiation.

In the case of Yang-Mills theory with the linear gauge fixing fermion
(\ref{psi}), the functional
$\Gamma[\phi_c,\phi^*_c]$ coincides with the reduced functional $\hat\Gamma$
introduced in \cite{Kluberg-Stern:1975rs,Zinn-Justin:1974mc}.
Hence, $\Gamma[\phi_c,\phi^*_c]$ can be
considered to be the generalization of this functional to the case of
generic gauge
theories with possibly non linear gauge fixing.

Let us compute the dependence of $\Gamma$ on the gauge
parameters: $\partial_{\xi^\alpha}\G=\partial_{\xi^\alpha}\G^\prime
=\partial_{\xi^\alpha}\G_{\rm
gf}|-\int d^nx\ \frac{\delta^R\G_{\rm
gf}}{\delta\tilde\phi^*_A}|\frac{\delta^L
\partial_{\xi^\alpha}\Psi}{\delta\phi^A}$,
where $|$ means that one has to substitute $\tilde\phi^*$ by
$\phi^*_c-\frac{\partial\Psi}{\partial \phi_c}$.
Using (\ref{6}) and the fact that the transformation is canonical,
it follows that
\bea
\partial_{\xi^\alpha}\G=\partial_{\xi^\alpha}\G^\prime=
(\Gamma_{\rm gf},{F_{\rm gf}}_\alpha)_{\phi_c,\tilde\phi^*}\nonumber\\
=(\Gamma^\prime,F_{\alpha})_{\phi_c,\phi^*_c},
\eea
where ${F_{\rm
      gf}}_\alpha[\phi_c,\tilde\phi^*_c]=L_{\alpha}\circ\G_{\rm gf}
+\partial_{\xi^\alpha}\Psi$ is of order at least $\hbar$, and
$F_{\alpha}[\phi_c,\phi^*_c]={F_{\rm
      gf}}_\alpha[\phi_c,\phi^*_c-\frac{\partial\Psi}{\partial
    \phi_c}]$.
{}From its classical limit, it is also clear that
$\frac{\delta\Gamma}{\delta y^\Delta_c}=O(\hbar)$, where
$y^\Delta_c\equiv(B^a_c,B^*_a,\bar C^a_c,\bar {C^*_a}_c)$ denote the
fields and antifields of the non minimal sector.
Thus, the functional
$\G[\phi_c,\phi^*_c]$ is independent, to order
$0$ in $\hbar$, of the gauge parameters and of the fields and
antifields of the non minimal sector.

\section*{Acknowledgments}

The authors acknowledge the hospitality of the Erwin Schr\"odinger
International Institute for Mathematical Physics in Vienna, where this
collaboration has been started, of the Max-Planck-Institut f\"{u}r Physik
(Werner-Heisenberg-Institut) in Munich and of the Department of
Theoretical Physics of the University of Valencia, where part of the work has
been done.  P.A.G.~wants to thank D.~Maison for long and useful
discussions. 

G.B.~is supported in part by the ``Actions de
Recherche Concert{\'e}es" of the ``Direction de la Recherche
Scientifique - Communaut{\'e} Fran{\c c}aise de Belgique", by
IISN - Belgium (convention 4.4505.86) and by
Proyectos FONDECYT 1970151 and 7960001 (Chile).
P.A.G.~is supported by NSF Grant No.\ PHY-9722083.

\newpage

\section{Appendix}
\setcounter{equation}{0}
\def\theequation{A.\arabic{equation}}

\begin{lemma}\label{00}
The insertion of a BRST exact local functional $(S_{\rm gf},\Xi)$ is equal to
$(\Gamma_{\rm gf},\cdot)$ applied to a local insertion, up to a local
insertion of higher order in $\hbar$.
\beq
(S_{\rm gf},\Xi)\circ\G_{\rm gf}=(\Gamma_{\rm gf},\Xi^Q\circ \G_{\rm
gf})
+\hbar I\circ\G_{\rm gf},\label{A1}
\eeq
where
$\Xi^Q=\Xi+O(\hbar)$ and
$I$ are local functionals.
\end{lemma}
Indeed, if $S^\infty$ is the sum of $S_{\rm gf}$ and the
BRST finite breaking local counterterms needed to achieve  (\ref{-1}),
the action $S_\rho=S^\infty+\Xi\rho$ satisfies
$\frac{1}{2}(S_\rho,S_\rho)=(S_{\rm gf},\Xi)\rho+O(\hbar)$, with
$\rho$ a Grassmann odd constant in ghost number $1$.
Applying the quantum
action principle, we get
$\frac{1}{2}(\G_\rho,\G_\rho)=\Delta(\rho)\circ\Gamma_\rho$. Putting
$\rho$ to zero, it follows from (\ref{-1}) that the local functional
$\Delta(0)=0$, so that
$\Delta(\rho)=\Delta^\prime(\rho)\rho$. Differentiation of the
previous equation with respect to $\rho$ and putting $\rho$ to zero
then implies
$(\Gamma_{\rm gf},\Xi^Q\circ\Gamma_{\rm gf})
= \Delta^\prime(0)\circ\G_{\rm gf}$, for some local functional
$\Xi^Q=\Xi+O(\hbar)$.
At tree level, this equation implies that $\Delta^\prime(0)=(S_{\rm
    gf},\Xi)+O(\hbar)$, which gives the result.
\bigskip

\noindent {\bf Proof of (\ref{2.1})~:}

It follows that $\partial_{\xi^\alpha}
\Gamma_{\rm gf}=(\Gamma_{\rm gf},[-\partial_{\xi^\alpha}\Psi]^Q\circ
\Gamma_{\rm gf})+\hbar K^\prime_\alpha\circ\Gamma_{\rm gf}$.
Applying $(\G_{\rm gf},\cdot)$
using (\ref{-1}), we get
to lowest order in $\hbar$ the consistency condition
$(S_{\rm gf},{K^\prime_\alpha }_0)=0$,
so that (\ref{coh}) implies ${K^\prime_{\alpha} }_0=-{\rho^i_a}_1
\partial_{g^i}S_{\rm gf}-(S_{\rm gf},{N_\alpha}_1)$. Using the quantum
action principle under the form $[\partial_{g^i}S_{\rm gf}]\circ\G_{\rm gf}
=\partial_{g^i}\G_{\rm gf}+\hbar I_i\circ\G_{\rm gf}$, for a local
insertion $I_i\circ\G$ and equation (\ref{A1}) again, we get
$\partial_{\xi^\alpha}\G_{\rm gf}-([
\partial_{\xi^\alpha}\Psi^Q+\hbar {N_\alpha}_1^Q]\circ
\Gamma_{\rm gf},\G_{\rm gf})+
\hbar{\rho^i_\alpha}_1\partial_{g^i}\G_{\rm gf}=\hbar^2
K^{\prime\prime}_\alpha\circ\G_{\rm gf}$,
and the reasoning can be pushed to higher orders.

\bigskip

\noindent {\bf Proof of (\ref{2.2})~:}

We introduce Grassmann odd ghost number $1$ parameters
$\lambda^\alpha$ and define $S^e=S^\infty +\lambda^\alpha
\partial_{\xi^\alpha}\Psi$. Using (\ref{5})
and
$\lambda^\alpha\lambda^\beta\frac{\partial^2\Psi}
{\partial\xi^\alpha\xi^\beta}=0$, it follows that
\beq
\frac{1}{2}(S^e,S^e)-\lambda^\alpha
D_\alpha S^e=\frac{1}{2}(\lambda^\alpha\partial_{\xi^\alpha}\Psi,
\lambda^\beta\partial_{\xi^\beta}\Psi)+O(\hbar),\label{Aex}
\eeq
where $O(\hbar)$ is
a local functional of order at least $\hbar$.
Applying the quantum action principle, it follows that
$\frac{1}{2}(\G^e,\G^e)-\lambda^\alpha
D_\alpha\G^e= \frac{1}{2}(\lambda^\alpha\partial_{\xi^\alpha}\Psi,
\lambda^\beta\partial_{\xi^\beta}\Psi)\circ\G^e+\hbar A\circ\G^e$.
Putting $\lambda^\alpha$ to zero and using (\ref{-1}), it follows that
$A=\lambda^\alpha A_\alpha$.
Differentiation with respect to $\lambda^\alpha$ and putting
$\lambda^\alpha$ to zero gives
$(\partial_{\xi^\alpha}\Psi^Q\circ \G_{\rm gf},
\G_{\rm gf})-D_\alpha\G_{\rm gf}=\hbar
A_{\alpha}(0)\circ\G_{\rm gf}$. Using (\ref{2.1}), we deduce that
$A_{\alpha}(0)\circ\G_{\rm gf}=(L^\prime_{\alpha}\circ \G_{\rm gf}
,\G_{\rm gf})$, where
$\hbar L^\prime_{\alpha}\circ \G_{\rm gf}=\partial_{\xi^\alpha}\Psi^Q\circ
\G_{\rm gf}+L_{\alpha}\circ
\G_{\rm gf}$. If we now add to $S^e$ the countertern $-\lambda^\alpha
\hbar {L^\prime_\alpha}_0$, we can absorb the lowest order contribution
$A_{\alpha}(\lambda=0)$ up to terms of second order in $\hbar$ or of
first order in $\hbar$ and of second order in $\lambda^\alpha$. For
the new $\G^e$, we end up with
$\frac{1}{2}(\G^e,\G^e)-\lambda^\alpha
D_\alpha\G^e=[\frac{1}{2}\lambda^\alpha\lambda^\beta
B_{[\alpha\beta]}(\lambda)+
\hbar^2 \lambda^\alpha
A^\prime_\alpha(0)]\circ\G^e$, where $B_{[\alpha\beta]}(\lambda)
= (\lambda^\alpha\partial_{\xi^\alpha}\Psi,
\lambda^\beta\partial_{\xi^\beta}\Psi)+O(\hbar)$.
Differentiation with respect to $\lambda^\alpha$ and
putting $\lambda^\alpha$ to zero now gives $(K_{\alpha}\circ\G_{\rm
gf},\G_{\rm gf})=\hbar^2
A^\prime_\alpha(0)\circ \G_{\rm gf}$. At order $0$ in $\hbar$, we get
${K_{\alpha}}_0=\mu^i_\alpha\partial_{g^i}S_{\rm gf}+(S_{\rm
gf},M_\alpha)$. But then
$K_{\alpha}\circ\G_{\rm gf}=\mu^i_\alpha\partial_{g^i}\G_{\rm gf}
+(\G_{\rm gf},M_\alpha^Q\circ \G_{\rm gf})+\hbar K^\prime _{\alpha}
\circ\G_{\rm gf}$ and we can forget about the first two terms, because
they are annihilated by $(\cdot,\G_{\rm gf})$. In the same way we can
get rid of the order $\hbar$ contribution and assume that
$\hbar^2(K^{\prime\prime}_{\alpha}\circ\G_{\rm
gf},\G_{\rm gf})=\hbar^2
A^\prime_\alpha(0)\circ \G_{\rm gf}$, which implies that
the lowest order contribution to $A^\prime_\alpha(0)$ can be absorbed
by adding suitable counterterm proportional to $\lambda^\alpha$ and of
order $\hbar^2$. Going on in the same way, one can achieve:
\beq
\frac{1}{2}(\G^e,\G^e)-\lambda^\alpha
D_\alpha\G^e
=\frac{1}{2}\lambda^\alpha\lambda^\beta K_{[\alpha\beta]}(\lambda)
\circ\G^e,
\eeq
where $K_{[\alpha\beta]}(\lambda)=(\lambda^\alpha\partial_{\xi^\alpha}\Psi,
\lambda^\beta\partial_{\xi^\beta}\Psi)+O(\hbar)$.

Acting with $\lambda^\gamma
D_\gamma$ on this equation, and using the same equation again, together with
$((\G^e,\G^e),\G^e)=0$, we find
$(-\frac{1}{2}\lambda^\alpha\lambda^\beta K_{[\alpha\beta]}(\lambda)
\circ\G^e,\G^e)-\frac{1}{2}\lambda^\alpha\lambda^\beta[D_{\alpha},D_\beta]
\G^e=\frac{1}{2}\lambda^\gamma\lambda^\alpha\lambda^\beta D_\gamma
[K_{[\alpha\beta]}(\lambda)
\circ\G^e]$. Differentiating with respect to $\lambda^\alpha$ and
$\lambda^\beta$ and putting $\lambda$ to zero gives (\ref{2.2}).

Note that (\ref{5}) is equivalent to
$\lambda^\alpha\partial_{\xi^\alpha} S_{\rm gf}=(S_{\rm gf},
\lambda^\alpha\partial_{\xi^\alpha}\Psi)$, which implies in particular that
$((\lambda^\alpha\partial_{\xi^\alpha}\Psi,
\lambda^\beta\partial_{\xi^\beta}\Psi),
S_{\rm gf})=0$, so that the right hand side of (\ref{2.2}) starts
indeed at order $\hbar$, as does the left hand side.

\bigskip

\begin {lemma}\label{l1}
The quantum analog of the classical condition (\ref{red})
on the couplings $g^i$ to be non redundant is
\beq
\mu^i\partial_{g^i}\G_{\rm gf}=(\G_{\rm gf},
\Xi^\prime\circ\G_{\rm gf})\Longrightarrow
\mu^i=0=(\G_{\rm gf},\Xi^\prime\circ\G_{\rm gf}),
\eeq
if $\Xi^\prime\circ\G_{\rm gf}$ is a local insertion.
\end{lemma}
Indeed, at tree level and
order $0$ in $\hbar$, we deduce because of (\ref{red}) that
$\mu^i_0=0=(S_{\rm gf},\Xi^\prime_0)$. It follows that
$\Xi^\prime_0=\lambda^i_0\partial_{g^i}S_{\rm gf}+(S_{\rm
    gf},\Theta_0)$, which implies
$\Xi^\prime\circ\G_{\rm gf}=\lambda^i_0\partial_{g^i}\G_{\rm gf}+(\G_{\rm gf},
\Theta_0\circ\G_{\rm gf})+\hbar\Xi^{\prime\prime}\circ\G_{\rm gf}$.
Using (\ref{-1}) and
$\mu^i_0=0$,
we have $\sum_{n\geq
    1}\hbar^n\mu^i_n\partial_{g^i}\G_{\rm gf}
=(\G,\hbar\Xi^{\prime\prime}\circ\G_{\rm gf})$. We now can
factorize $\hbar$ and the reasoning can be pushed to higher orders.

\bigskip

\noindent {\bf Proof of (\ref{2})~:}

The quantum action principle implies
$\mu\partial_\mu \Gamma_{\rm gf}+\hbar I\circ \Gamma_{\rm gf}=0$. Applying
$(\Gamma_{\rm gf},\cdot)$ and using (\ref{-1}), we get to lowest order
$(S_{\rm
gf},I_0)=0$. Using stability, this implies $I_0=\bar \beta_1^i\partial_{g^i}
S_{\rm gf}+(C_1,S_{\rm gf})$. By the same reasoning than above,
this implies $\mu\partial_\mu \Gamma_{\rm gf}+\hbar\bar
\beta_1^i\partial_{g^i}\G_{\rm gf}+\hbar(C_1\circ\G_{\rm gf},\G_{\rm gf})
+\hbar^2I^\prime\circ\G_{\rm gf}=0$, and
the reasoning can be pushed to higher orders.

\bigskip

\noindent {\bf Proof of $[D,D_\alpha]^i=0$~:}

Consider the parameters $\xi^{\bar\alpha}=(\xi^\alpha,\mu)$,
the ghosts $\lambda^{\bar \alpha}=(\lambda^\alpha,\Lambda)$
and the differentials
$D_{\bar\alpha}=(D_{\alpha},D)$. It then follows that (\ref{Aex})
holds for the same $S^e$ but with $\lambda^\alpha D_{\alpha}$ replaced
by $\lambda^{\bar \alpha} D_{\bar \alpha}$. The proof that $[
D_{\bar \alpha}, D_{\bar \beta}]=0$ then proceeds exactly as before
and includes the result we need, i.e., $[D,D_{\alpha}]=0$.

\vfill
\pagebreak

\end{document}